\documentclass[a4paper,11pt]{article}
\usepackage{pos}

\usepackage{acronym}

\usepackage{epsf,graphicx,scalefnt,rotating,enumitem,fancyhdr}
\usepackage{amsmath}

\usepackage{amssymb}
\usepackage{dsfont}
\usepackage[english]{babel}
\usepackage{multirow}
\usepackage[final]{showkeys}
\usepackage{slashed}
\usepackage{filemod}

\newcounter{notecount}

\newcommand{\lmut}{L_{\mu t}}

\newcommand{\Zchiring}{\mathring{Z}_\chi}

\newcommand{\ctr}{T_\text{R}}
\newcommand{\tr}{\ctr}
\newcommand{\nc}{n_\text{c}}

\newcommand{\nf}{n_\text{f}}

\newcommand{\ren}{\text{\abbrev{R}}}
\newcommand{\bare}{\text{\abbrev{B}}}
\newcommand{\citere}[1]{Ref.~\cite{#1}}
\newcommand{\citeres}[1]{Refs.~\cite{#1}}

\newcommand{\abbrev}[1]{{\scalefont{.9}#1}}
\newcommand{\EulerGamma}{\gamma_\text{E}}
\newcommand{\ep}{\epsilon}
\newcommand{\api}{a_s}

\newcommand{\dd}{\mathrm{d}}

\newcommand{\deriv}[3]{\frac{\partial\ifthenelse{\equal{#1}{}}{}{^{#1}}%
    #2}{\partial #3\ifthenelse{\equal{#1}{}}{}{^{#1}}}}
\newcommand{\dderiv}[3]{\frac{\dd\ifthenelse{\equal{#1}{}}{}{^{#1}}%
    #2}{\dd #3\ifthenelse{\equal{#1}{}}{}{^{#1}}}}

\newcommand{\msbar}{\ensuremath{\overline{\mbox{\abbrev{MS}}}}}
\newcommand{\drbar}{\ensuremath{\overline{\mbox{\abbrev{DR}}}}}

\newcommand{\tcalo}{\tilde{\calo}}

\newcommand{\calo}{\mathcal{O}}
\newcommand{\cale}{\mathcal{O}}
\newcommand{\PP}{\calo\calo}
\newcommand{\PE}{\calo E}
\newcommand{\EP}{E\calo}
\newcommand{\EE}{EE}

\newcommand{\myacrodef}[3]{\acrodef{#2}{#3}\newcommand{#1}{\ac{#2}}}

\acused{HKL}
\newcommand{\qcd}{\abbrev{QCD}}
\acused{QCD} 

\allowdisplaybreaks

\usepackage{array}
\usepackage[capitalize]{cleveref}

\title{The gradient flow formulation of the electroweak Hamiltonian}

\author*[a,b]{Fabian Lange}

\affiliation[a]{Institut f\"ur Theoretische Teilchenphysik, Karlsruhe Institute of Technology (KIT), \\
  Wolfgang-Gaede-Stra\ss{}e 1, 76128 Karlsruhe, Germany}

\affiliation[b]{Institut f\"ur Astroteilchenphysik, Karlsruhe Institute of Technology (KIT), \\
  Hermann-von-Helmholtz-Platz 1, 76344 Eggenstein-Leopoldshafen, Germany}

\emailAdd{fabian.lange@kit.edu}

\abstract{Flavor observables are usually computed with the help of the electroweak Hamiltonian which separates the short-distance from the long-distance regime.
The Wilson coefficients are calculated perturbatively, while matrix elements of the operators require non-perturbative treatment for many processes, e.g.\ through lattice simulations.
The resulting necessity to compute the transformation between the different renormalization schemes in the two calculations constitutes an important source of uncertainties.
An elegant solution to this problem is provided by the gradient-flow formalism, already widely used in lattice simulations, because its composite operators do not require renormalization.
In this contribution we report on the construction of the electroweak Hamiltonian in the gradient-flow formalism through NNLO in QCD.}

\FullConference{%
  Loops and Legs in Quantum Field Theory - LL2022,\\
  25-30 April, 2022\\
  Ettal, Germany
}

\begin{document}
\maketitle

\myacrodef{\sftx}{SFTX}{small-flow-time expansion}
\myacrodef{\cmm}{CMM}{Chetyrkin-Misiak-M\"unz}
\myacrodef{\wrt}{w.r.t.}{with respect to}
\myacrodef{\vpf}{VPF}{vacuum polarization function}
\myacrodef{\vev}{VEV}{vacuum expectation value}
\myacrodef{\rg}{RG}{renormalization group}
\myacrodef{\gff}{GFF}{gradient-flow formalism}
\myacrodef{\ope}{OPE}{Operator Product Expansion}
\myacrodef{\ckm}{CKM}{Cabbibo-Kobayashi-Maskawa}
\myacrodef{\lhc}{LHC}{Large Hadron Collider}
\myacrodef{\uv}{UV}{ultraviolet} \myacrodef{\lo}{LO}{leading order}
\myacrodef{\nlo}{NLO}{next-to-leading order}
\myacrodef{\nnlo}{NNLO}{next-to-next-to-leading order}
\myacrodef{\llog}{LL}{leading logarithmic}
\myacrodef{\nnll}{NNLL}{next-to-next-to-leading logarithmic}
\myacrodef{\pdf}{PDF}{parton density function}
\myacrodef{\sm}{SM}{Standard Model}
\myacrodef{\bsm}{BSM}{beyond-the-\ac{SM}}
\myacrodef{\mssm}{MSSM}{Minimal Supersymmetric \ac{SM}}
\myacrodef{\susy}{SUSY}{Supersymmetry}
\myacrodef{\dreg}{DREG}{Dimensional Regularization}
\myacrodef{\dred}{DRED}{Dimensional Reduction}
\myacrodef{\emt}{EMT}{energy-momentum tensor}

\section{Introduction}

The effective electroweak Hamiltonian is often used to compute predictions for flavor observables because it separates the short-distance from the long-distance contributions.
For many processes the latter are non-perturbative.
The Wilson coefficients are then usually computed within perturbation theory, while the non-perturbative matrix elements of the effective operators are computed via lattice simulations or sum rules.
However, to combine both ingredients to a physical prediction, the different schemes used in the individual calculations have to be matched, which constitutes an important source of uncertainties.

The \gff~\cite{Narayanan:2006rf,Luscher:2009eq,Luscher:2010iy} offers a promising solution to this problem because flowed operators are \uv-finite.
Thus, the flowed operators can be used both in perturbative calculations as well as in lattice simulations.
The matching between the regular and the flowed operators is perturbative and can be absorbed into flow-time dependent Wilson coefficients.
This strategy has already successfully been applied to the energy-momentum tensor of \qcd{} through \nnlo{}~\cite{Suzuki:2013gza,Makino:2014taa,Harlander:2018zpi}, which led to competitive thermodynamical results, see
e.g.\ \citeres{Iritani:2018idk,Taniguchi:2020mgg,Shirogane:2020muc,Suzuki:2021tlr}.
Furthermore, the matching matrix has also been calculated for quark-dipole operators through \nlo{} \qcd{}~\cite{Rizik:2020naq,Mereghetti:2021nkt} and for hadronic vacuum polarization through \nnlo\ \qcd~\cite{Harlander:2020duo}.

For the effective electroweak Hamiltonian the matching matrix for the current-current operators has been calculated at \nlo\ \qcd\ in the \drbar\ scheme in \citere{Suzuki:2020zue}.
In this contribution we report on our recent calculation~\cite{Harlander:2021jcq,Harlander:2022tgk} of the same matching matrix through \nnlo\ in the basis defined in \citere{Chetyrkin:1997gb} which allows us to adopt the \msbar\ scheme with a fully anti-commuting $\gamma_5$.
Our result could directly be applied to predict $K$- or $B$-mixing parameters on the basis of the \gff{} once the corresponding matrix elements from lattice simulations become available.


\section{Operator basis}

Schematically the effective electroweak Hamiltonian can be written as
\begin{equation}
  \label{eq:hamil} \mathcal{H}_\mathrm{eff} =
  - \left(\frac{4G_\mathrm{F}}{\sqrt{2}}\right)^x V_\mathrm{CKM} \, \sum_n
  C_n \calo_n
\end{equation}
where $G_\mathrm{F}$ denotes the Fermi constant raised to some power $x$, $V_\mathrm{CKM}$ comprises
the relevant elements of the \ckm\ matrix, and $C_n$ are the Wilson coefficients.
As a first step we focus on the current-current operators
\begin{equation}
  \label{eq:phys} \begin{split}
  \calo_1 &=
  - \left(\bar\psi_{1} \gamma_\mu^{\mathrm{L}}
  T^a \psi_{2}\right) \left(\bar\psi_{3} \gamma_\mu^{\mathrm{L}}
  T^a \psi_{4}\right) ,\\ \calo_2
  &= \left(\bar\psi_{1} \gamma_\mu^{\mathrm{L}} \psi_{2}\right)
  \left(\bar\psi_{3} \gamma_\mu^{\mathrm{L}} \psi_{4}\right) ,
  \end{split}
\end{equation}
where we choose the basis of \citere{Chetyrkin:1997gb}.
Furthermore, we adopt the Euclidean metric and use the short-hand notation
\begin{equation}\label{eq:basis:hunt}
  \begin{split}
    \gamma_\mu^{\mathrm{L}}= \gamma_\mu\frac{1-\gamma_5}{2}
  \end{split}
\end{equation}
to project onto the left-handed components of the spinors.
Our convention for the color generators is
\begin{equation}\label{eq:easy}
  \begin{split}
    [T^a,T^b] = f^{abc}T^c\,,\quad \text{Tr}(T^aT^b) = -\tr\delta^{ab}\,,
  \end{split}
\end{equation}
with the real and totally anti-symmetric structure constants $f^{abc}$ and the trace normalization $\tr$.
In addition to the physical operators in \cref{eq:phys}, loop corrections in dimensional regularization with $D=4-2\ep$ introduce contributions which have to be attributed to so-called evanescent operators.
Even though they vanish for $D = 4$, they mix with the physical operators at higher orders in perturbation theory~\cite{Buras:1989xd}.
We again follow \citere{Chetyrkin:1997gb} and choose
\begin{equation}
  \label{eq:basis:feed} \begin{split}
    \cale_1^{(1)} &=
  - \left(\bar\psi_{1} \gamma_{\mu\nu\rho}^{\mathrm{L}}
  T^a \psi_{2}\right) \left(\bar\psi_{3} \gamma_{\mu\nu\rho}^{\mathrm{L}}
  T^a \psi_{4}\right) - 16 \calo_1 ,\\ \cale_2^{(1)}
  &= \left(\bar\psi_{1} \gamma_{\mu\nu\rho}^{\mathrm{L}}
  \psi_{2}\right) \left(\bar\psi_{3}
  \gamma_{\mu\nu\rho}^{\mathrm{L}} \psi_{4}\right)
  - 16 \calo_2\,,\\
  \cale_1^{(2)} &=
  - \left(\bar\psi_{1} \gamma_{\mu\nu\rho\sigma\tau}^{\mathrm{L}}
  T^a \psi_{2}\right) \left(\bar\psi_{3}
  \gamma_{\mu\nu\rho\sigma\tau}^{\mathrm{L}}
  T^a \psi_{4}\right) - 20 \calo_1^{(1)} - 256 \calo_1 ,\\
  \cale_2^{(2)}
  &= \left(\bar\psi_{1} \gamma_{\mu\nu\rho\sigma\tau}^{\mathrm{L}}
  \psi_{2}\right) \left(\bar\psi_{3} \gamma_{\mu\nu\rho\sigma\tau}^{\mathrm{L}}
  \psi_{4}\right) - 20 \calo_2^{(1)} - 256 \calo_2 .
  \end{split}
\end{equation}
as evanescent operators, where $\gamma_{\rho\mu_1 \cdots
  \mu_n}^{\mathrm{L}} \equiv \gamma_\rho^{\mathrm{L}}\gamma_{\mu_1}
\cdots \gamma_{\mu_n}$. We will refer to the basis defined by
\cref{eq:phys,eq:basis:feed} as the \cmm-basis in what follows.


\section{Flowed operators}

In the \gff, one introduces flowed gluon and quark fields
$B^a_\mu=B^a_\mu(t)$ and $\chi=\chi(t)$ as solutions of the
flow equations~\cite{Luscher:2010iy,Luscher:2013cpa}
\begin{equation}
  \begin{split}
    \partial_t B^a_\mu &= \mathcal{D}^{ab}_\nu G^b_{\nu\mu} + \kappa
    \mathcal{D}^{ab}_\mu \partial_\nu B^b_\nu\,,\\ \partial_t \chi
    &= \Delta \chi - \kappa \partial_\mu B^a_\mu T^a \chi\,,\\ \partial_t
    \bar \chi &= \bar \chi \overleftarrow \Delta + \kappa \bar
    \chi \partial_\mu B^a_\mu T^a\,,
    \label{eq:flow}
  \end{split}
\end{equation}
with the initial conditions
\begin{equation}
  \begin{split}
    B^a_\mu (t=0) = A^a_\mu\,,\qquad \chi (t=0)= \psi\,,
    \label{eq:bound}
  \end{split}
\end{equation}
where $A^a_\mu$ and $\psi$ are the regular gluon and quark fields,
respectively, and
\begin{equation}\label{eq:dleftright}
  \begin{split}
    \mathcal{D}^{ab}_\mu &= \delta^{ab}\partial_\mu - f^{abc}
    B_\mu^c\,,\\
    G_{\mu\nu}^a &= \partial_\mu B_\nu^a -
    \partial_\nu B_\mu^a + f^{abc}B_\mu^bB_\nu^c\,, \\
    \Delta &= (\partial_\mu + B^a_\mu T^a)^2\,.
  \end{split}
\end{equation}
The parameter $\kappa$ is arbitrary and drops out of physical
quantities; we will set $\kappa=1$ in our calculation, because this
choice reduces the size of the intermediate algebraic expressions.

Our practical implementation of the \gff\ in perturbation theory follows
the strategy developed in \citere{Luscher:2011bx} and further detailed
in \citere{Artz:2019bpr}.
The \qcd{} propagators are generalized by multiplying them with flow-time-dependent exponential functions.
The flow equations are introduced on the Lagrangian level with the help of Lagrange multiplier fields.
This leads to propagators directed towards increasing flow time, the so-called ``flow lines'', and integrations over the flow times of new ``flowed vertices''.

These exponential functions regulate some of the \uv{}-divergencies so that the flowed gluon field $B^a_\mu$ does not require renormalization~\cite{Luscher:2010iy,Luscher:2011bx}.
The flowed quark fields $\chi$, on the other hand, have to be renormalized~\cite{Luscher:2013cpa}.
We choose the ringed scheme for which the renormalization constant $\Zchiring$ is defined by the all-order condition
\begin{equation}\label{eq:zchidef}
  \begin{split}
    \Zchiring\langle \bar\chi\overleftrightarrow{\mathcal{\slashed{D}}}
    \chi\rangle_0\bigg|_{m=0} &\equiv -\frac{2\nc}{(4\pi t)^2}\,,\\[.5em]
    \overleftrightarrow{\mathcal{D}}_\mu = \partial_\mu
    -&\overleftarrow{\partial}\!_\mu + 2B_\mu^a T^a\,,
  \end{split}
\end{equation}
where $\langle\cdot\rangle_0$ denotes the \vev~\cite{Makino:2014taa}.
It is known through \nnlo{}~\cite{Artz:2019bpr}.

Furthermore, it was shown that composite operators constructed from flowed fields are \uv\ finite after the renormalization of the strong coupling, the quark masses, and the flowed fields~\cite{Luscher:2011bx}.
We simply define the flowed operators by replacing the spinors $\psi_i$ by renormalized flowed spinors $\Zchiring^{1/2}\chi_i$ in the regular operators, i.e.
\begin{equation}\label{eq:flowed:craw}
  \begin{split}
      \tcalo_1 &=
  - \Zchiring^2\left(\bar\chi_{1} \gamma_\mu^{\mathrm{L}}
  T^a \chi_{2}\right) \left(\bar\chi_{3} \gamma_\mu^{\mathrm{L}}
  T^a \chi_{4}\right) ,\\
  \tcalo_2
  &= \Zchiring^2\left(\bar\chi_{1} \gamma_\mu^{\mathrm{L}} \chi_{2}\right)
  \left(\bar\chi_{3} \gamma_\mu^{\mathrm{L}} \chi_{4}\right)\,,
  \end{split}
\end{equation}
and analogously for the evanescent operators.
Since they are finite, one can treat them in four space-time dimensions, which also means that flowed evanescent operators vanish.
By keeping them in our calculation we can check this explicitly as a welcome consistency check on our results.
Let us stress that the regular evanescent operators are still needed.


\section{Small-flow-time expansion}

One can relate the flowed and the regular operators through the small-flow-time expansion $t\to 0$~\cite{Luscher:2011bx}
\begin{equation}\label{eq:esau}
  \begin{split}
      \begin{pmatrix}
        \tcalo(t) \\
        \tilde{E}(t)
  \end{pmatrix}
      \asymp \zeta^\bare(t) \begin{pmatrix} \calo \\ E
      \end{pmatrix}\,,
  \end{split}
\end{equation}
where we use the notation
\begin{equation}\label{eq:bohr}
  \begin{split}
    \calo &= (\calo_1,\calo_2)^\mathrm{T}\equiv
    (\calo_1^{(0)},\calo^{(0)}_2)^\mathrm{T}\,,\\
    E &= (\cale^{(1)}_1,\cale^{(1)}_2,\cale^{(2)}_1,\cale^{(2)}_2)^\mathrm{T}\,,
  \end{split}
\end{equation}
and analogously for the flowed operators. Here and in what follows, the
superscript ``$\bare$'' denotes a ``bare'' quantity which will undergo
renormalization.
Terms of $O(t)$ are neglected as indicated by the symbol $\asymp$.
We also adopt the block-notation of \cref{eq:esau} for matrices, e.g.\ for the renormalized matching matrix we write
\begin{equation}\label{eq:gong}
  \begin{split}
    \zeta(t) =
    \left(
    \begin{matrix}
      \zeta_{\PP}(t) & \zeta_{\PE}(t)\\
      \zeta_{\EP}(t) & \zeta_{\EE}(t)
    \end{matrix}
    \right) ,
  \end{split}
\end{equation}
where the $2\times2$-submatrix $\zeta_{\PP}$ concerns only the physical
operators.

While the flowed operators on the l.h.s.\ of \cref{eq:esau} are finite, the regular operators on the r.h.s.\ are divergent.
Hence, the bare matching matrix $\zeta^\bare(t)$ is divergent for $D\to4$ as well.
However, one may define renormalized operators
\begin{equation}
  \label{eq:EW:ren}
  \begin{pmatrix}
    \calo \\
     E
  \end{pmatrix}^\ren
  = Z
  \begin{pmatrix}
    \calo \\
     E
  \end{pmatrix}
  \equiv
  \begin{pmatrix}
    Z_{\PP} & Z_{\PE} \\
    Z_{\EP} & Z_{\EE}
  \end{pmatrix}
  \begin{pmatrix}
    \calo \\
    E
  \end{pmatrix}
\end{equation}
with the corresponding renormalization matrix $Z$ such that matrix elements of them are finite.
Usually, the blocks $Z_{\PP}$, $Z_{\PE}$, and $Z_{\EE}$ are defined in the $\msbar$ scheme.
On the other hand, the block $Z_{\EP}$ is finite and chosen such that physical matrix elements $\langle\cdot\rangle$ of evanescent operators vanish to all orders in perturbation theory~\cite{Buras:1989xd,Dugan:1990df,Herrlich:1994kh}, i.e.
\begin{equation}\label{eq:EW:amoy}
  \begin{split}
    \langle E^\ren\rangle &=
    Z_{\EP}\langle \calo\rangle
    + Z_{\EE}\langle E\rangle \overset{!}{=}O(\ep)\,.
  \end{split}
\end{equation}
By inserting \cref{eq:EW:ren} into \cref{eq:esau}, we can then define the renormalized matching matrix
\begin{equation}
  \label{eq:EW_SMT_ren}
  \zeta(t) = \zeta^\bare(t)Z^{-1} =
  \left(
  \begin{matrix}
    \zeta_{\PP}(t) & \zeta_{\PE}(t)\\
    \zeta_{\EP}(t) & \zeta_{\EE}(t)
  \end{matrix}
  \right) .
\end{equation}
Since $\langle \tilde{E}(t)\rangle = O(\ep)$, the
renormalization condition in \cref{eq:EW:amoy} is equivalent to
\begin{equation}\label{eq:junr}
  \begin{split}
    \zeta_{\EP}(t)=O(\ep)\,.
  \end{split}
\end{equation}


\section{Calculation of the matching matrix}

To compute the matching matrix $\zeta(t)$ we employ the method of projectors~\cite{Gorishnii:1983su,Gorishnii:1986gn}.
The projectors are matrix elements
\begin{equation}\label{eq:calculation:cion}
  \begin{split}
    P^{(i)}_j[X] = \langle0|X|i,j\rangle\bigg|_{p=m=0}\,,
  \end{split}
\end{equation}
with $i\in\{0,1,2\}$ and $j\in\{1,2\}$, such that
\begin{equation}\label{eq:deli}
  \begin{split}
    P^{(i)}_j[\calo^{(i')}_{j'}] = \delta_{ii'}\delta_{jj'}\,,
    \end{split}
\end{equation}
where we remind the reader of the unified notation for physical
and evanescent operators defined in \cref{eq:bohr}.  In general, the
projectors could also involve derivatives w.r.t.\ masses and/or external
momenta, but this is not necessary for the set of operators considered
here. By setting all external mass scales to zero in
\cref{eq:calculation:cion}, it is sufficient to satisfy \cref{eq:deli}
at tree-level, because all higher perturbative orders on the
l.h.s.\ vanish in dimensional regularization.

The external states $|i,j\rangle$ are defined to explicitly project onto left-handed spinors.
Together with an anti-commuting $\gamma_5$ this eliminates all $\gamma_5$'s from the traces at any order in the calculation~\cite{Chetyrkin:1997gb}.

\begin{figure}
  \begin{center}
    \begin{tabular}{ccc}
          \includegraphics[%
            width=.22\textwidth]%
                          {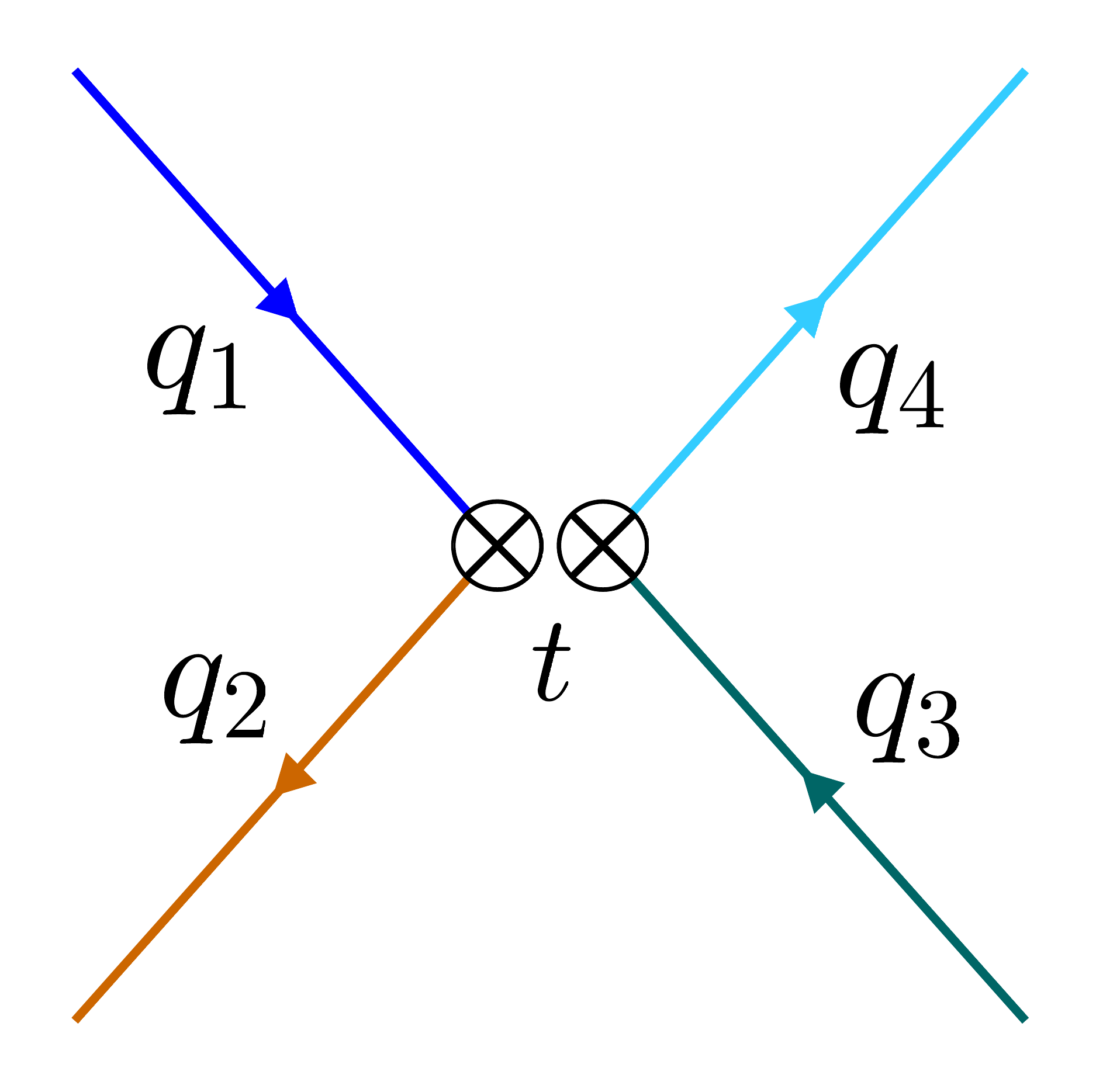} &
          \includegraphics[%
            width=.25\textwidth]%
                          {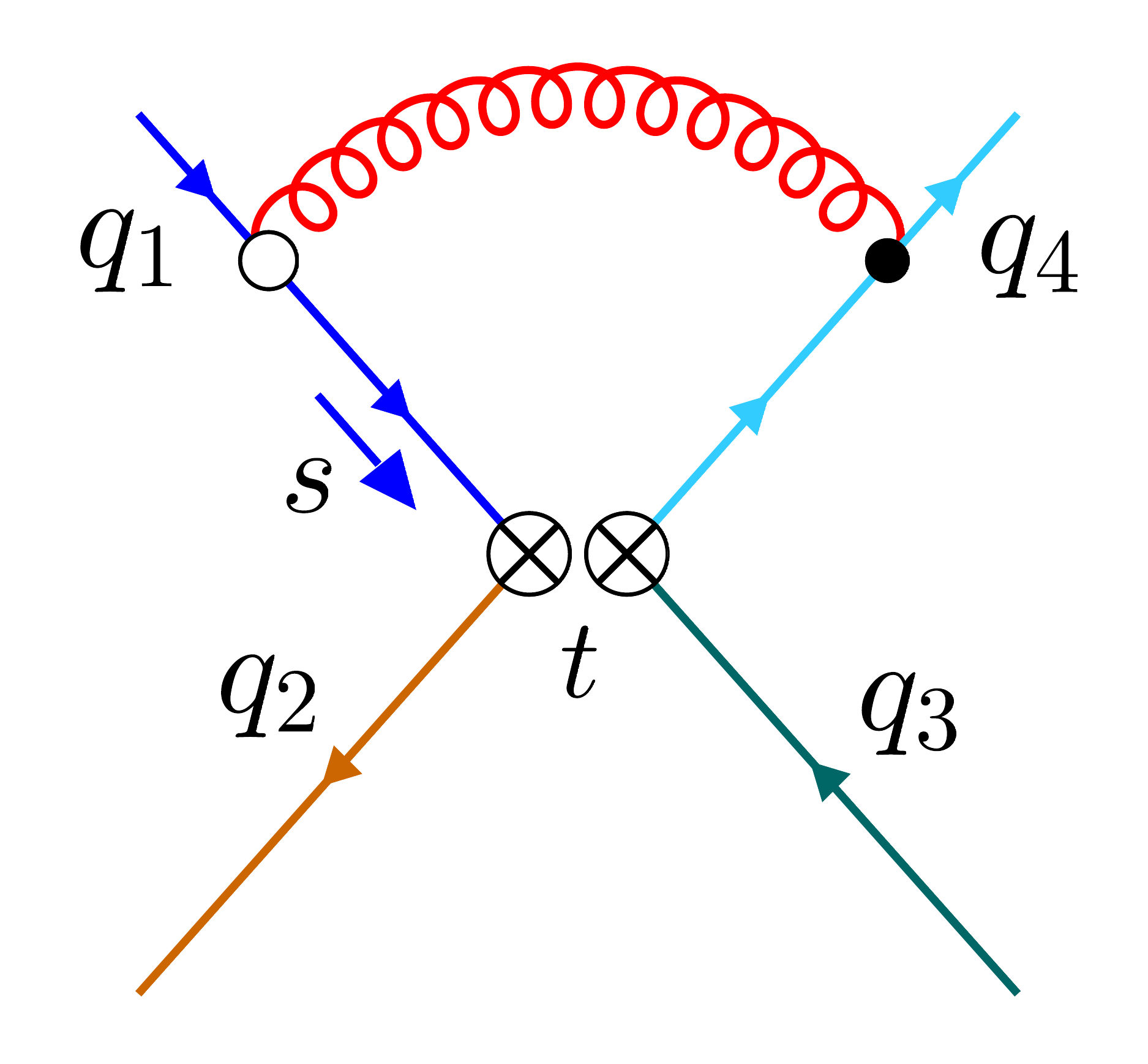} &
          \includegraphics[%
            width=.25\textwidth]%
                          {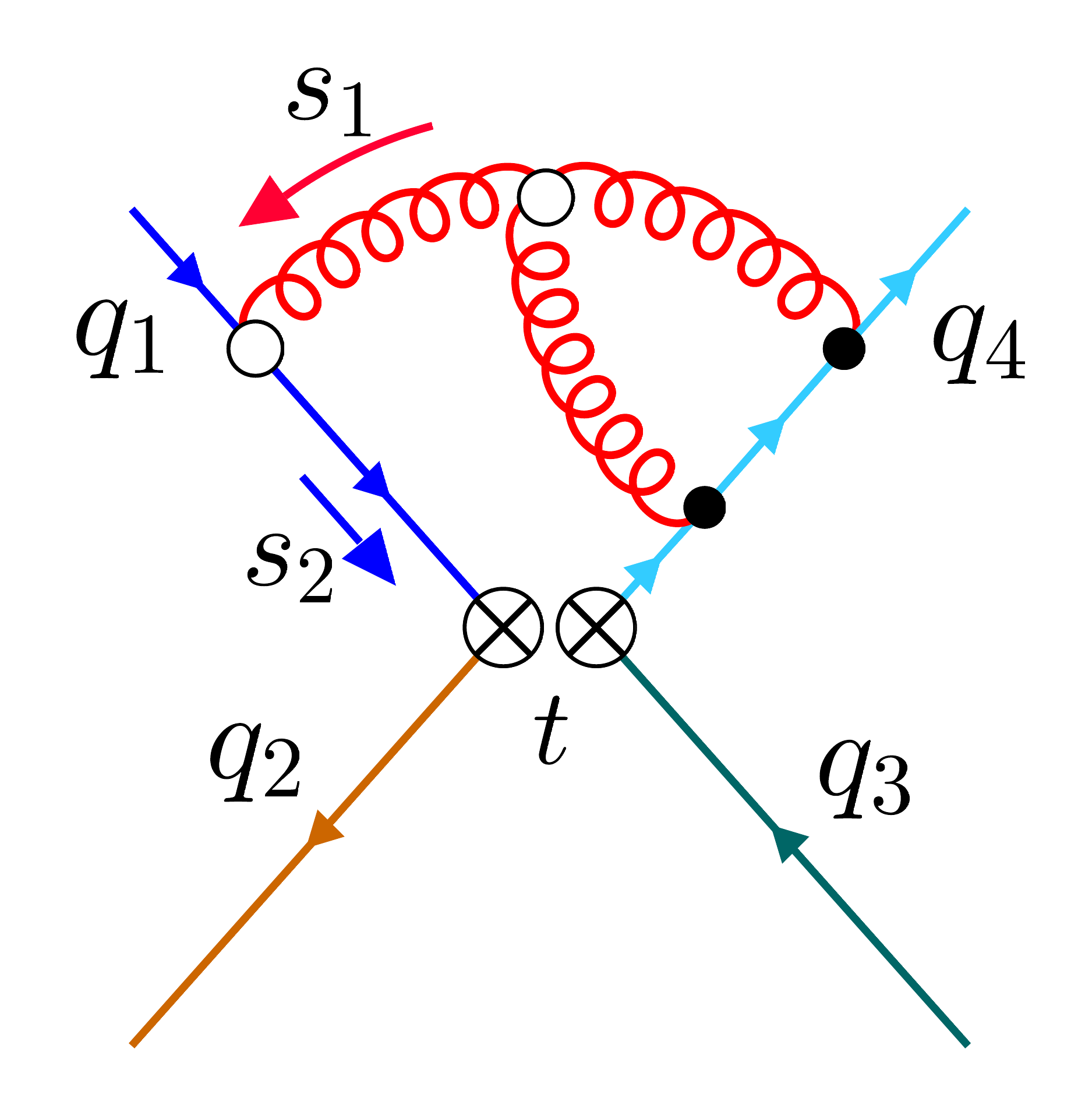}\\[-1em]
                            (a) & (b) & (c)
    \end{tabular}
      \caption[]{\label{fig:dias}\sloppy Sample diagrams contributing to the determination of the matching matrix $\zeta(t)$ at \lo, \nlo, and \nnlo\ \qcd. The circles denote ``flowed vertices'', lines with an arrow next to them denote ``flow lines'', and the label next to the arrow is a flow-time integration variable (see \citere{Artz:2019bpr} for details). The diagrams were produced with \texttt{FeynGame}~\cite{Harlander:2020cyh}.}
  \end{center}
\end{figure}

Applying the projectors to \cref{eq:esau} directly yields the bare matching matrix
\begin{equation}\label{eq:dory}
  \begin{split}
    \zeta^{\bare,(ii')}_{jj'}(t) = P^{(i')}_{j'}[\tcalo^{(i)}_j(t)] ,
  \end{split}
\end{equation}
where $\zeta^{(00)}_{jj'} \equiv (\zeta_{\PP})_{jj'}$.
By restricting the calculation to the case with four different quark flavors in the operators, all Feynman diagrams contributing to the r.h.s.\ of this equation are obtained by dressing the generic tree-level diagram in \cref{fig:dias}\,(a).
Sample diagrams are shown in \cref{fig:dias}\,(b) and (c).


%

%


The actual calculation of the diagrams is performed with the setup described in \citere{Artz:2019bpr} based on \texttt{q2e/exp}~\cite{Harlander:1998cmq,Seidensticker:1999bb}:
After generating the Feynman diagrams with \texttt{qgraf}~\cite{Nogueira:1991ex,Nogueira:2006pq}, we apply the projectors, perform the traces, and simplify the algebraic expressions within \texttt{FORM}~\cite{Vermaseren:2000nd,Kuipers:2012rf,vanRitbergen:1998pn}.
With the help of \texttt{Kira+FireFly}~\cite{Maierhofer:2017gsa,Klappert:2020nbg,Klappert:2019emp,Klappert:2020aqs}, the resulting Feynman integrals are reduced to the same master integrals as found in \citere{Harlander:2018zpi}.


\section{Results}

Performing the calculation and renormalization as described in the previous sections, our results for the physical components of the renormalized matching matrix through \nnlo{} in \qcd{} read~\cite{Harlander:2022tgk}
\begin{equation}\label{eq:EW:glob}
  \begin{split}
    (\zeta^{-1})_{11}(t) &= 1
    +\api\, \left(4.212+ \frac{1}{2} \lmut\right)
    + \api^2 \bigg[
    22.72-0.7218\, \nf
    +\lmut \left(16.45-0.7576\, \nf\right) \\
    & \qquad\qquad\qquad\qquad\qquad\qquad \ \ + \lmut^2 \left(\frac{17}{16}-\frac{1}{24}\, \nf\right)
    \bigg]\,,\\
    (\zeta^{-1})_{12}(t) &=
    \api\,\left(
    -\frac{5}{6}
    - \frac{1}{3}\lmut
    \right)
    + \api^2\,\bigg[
    -4.531 + 0.1576\,\nf
    + \lmut\,\left(-3.133 + \frac{5}{54}\,\nf\right) \\
    & \qquad\qquad\qquad\qquad\qquad \ \ + \lmut^2\,\left(-\frac{13}{24} + \frac{1}{36}\nf\right)
    \bigg]\,,
     \\
     (\zeta^{-1})_{21}(t) &=
     \api\,\left(-\frac{15}{4} - \frac{3}{2}\,\lmut\right)
     + \api^2\,\bigg[-23.20
     + 0.7091\,\nf
     + \lmut\,\left(-15.22
     + \frac{5}{12}\,\nf\right) \\
     & \qquad\qquad\qquad\qquad\qquad\quad + \lmut^2\,\left(-\frac{39}{16} + \frac{1}{8}\,\nf\right)
     \bigg]\,,\\
     (\zeta^{-1})_{22}(t) &=
     1
     + 3.712\,\api
     + \api^2\,\bigg[
     19.47
     - 0.4334\,\nf
     + \lmut\,\left(11.75 - 0.6187\,\nf\right)
     + \frac{1}{4}\,\lmut^2
     \bigg]\,,
  \end{split}
\end{equation}
with $\api=\alpha_s(\mu)/\pi$ and $\lmut=\ln 2\mu^2t + \EulerGamma$, where $\alpha_s$ is the strong coupling renormalized in the \msbar\ scheme with $\nf$ quark flavors, $\mu$ the renormalization scale, and $\EulerGamma=0.577\ldots$ Euler's constant. For the sake of compactness, we set $\nc=3$ and $\tr=\tfrac{1}{2}$, and replaced transcendental coefficients by floating-point numbers. Analytical coefficients for a general SU($\nc$) gauge group are included in an ancillary file of \citere{Harlander:2022tgk}.

The correctness of our result is supported by several observations:
First of all, the renormalized matching matrix is finite when we employ the literature expression for the renormalization matrix $Z$ from \citeres{Chetyrkin:1997gb,Gambino:2003zm,Gorbahn:2004my}.
In addition, \cref{eq:junr} is fulfilled with the same $Z$.
Furthermore, our result is gauge independent even though we performed the calculation in $R_\xi$ gauge.
The final non-trivial check is the basis transformation to the so-called non-mixing basis of \citere{Buras:2006gb}, which is defined such that the anomalous dimension matrix for the physical operators is diagonal through \nnlo{}.
In this basis the physical matching matrix $\zeta(t)$ between the \msbar\ renormalized and the flowed operators turns out to be diagonal.\footnote{An immediate comparison of this result to the \nlo\ expression of \citere{Suzuki:2020zue} is not possible, because the latter is obtained in the \drbar\ scheme.}
The result in the non-mixing basis can be found in \citere{Harlander:2022tgk}.



\section{The effective Hamiltonian in the gradient-flow formalism}

By inverting the small-flow-time expansion in \cref{eq:esau}, one can rewrite the effective electroweak Hamiltonian as
\begin{equation}
  \label{eq:hamilflow} \mathcal{H}_\mathrm{eff} \asymp
  - \left(\frac{4G_\mathrm{F}}{\sqrt{2}}\right)^x V_\mathrm{CKM} \, \sum_n
  \tilde{C}_n(t) \tcalo_n(t)
\end{equation}
to express it in terms of the flowed operators.
The flowed Wilson coefficients are given by
\begin{equation}\label{eq:arad}
  \begin{split}
    \tilde{C}_n(t) = \sum_{m}C_m^\ren\,\zeta^{-1}_{mn}(t)\,,
  \end{split}
\end{equation}
with $\zeta(t)\equiv \zeta_{\PP}(t)$ the physical part of the matching matrix,
and $C_n^\ren=\sum_{m}C_m(Z^{-1})_{mn}$ the renormalized regular Wilson
coefficients.
Since both the flowed operators and the flowed Wilson coefficients are individually finite without operator renormalization, they are also individually scheme and renormalization scale independent (up to higher orders in perturbation theory).
This is in stark contrast to $C^\ren$ and $\calo^\ren$ which depend on the renormalization scheme, including the treatment of $\gamma_5$ and the choice of evanescent operators.
\cref{eq:hamilflow} can thus be used both perturbatively and on the lattice without matching perturbative and lattice schemes.
However, on the perturbative side it is important to evaluate $C^\ren$ and $\zeta^{-1}(t)$ in the same renormalization scheme.

Re-expanding the r.h.s.\ of \cref{eq:arad}, directly gives the flowed Wilson coefficients to the known order of either $C^\ren$ or $\zeta^{-1}(t)$, whichever is lower.
For Kaon mixing, i.e.\ $|\Delta S| = 2$, the physical operator basis from \cref{eq:phys} reduces to just one operator due to a Fierz identity.
In this case, the \sm{} Wilson coefficient is known through \nlo{}~\cite{Buchalla:1995vs}, with two contributions known through \nnlo{}~\cite{Brod:2010mj,Brod:2011ty}.
Similarly, only one operator contributes to the mass difference of neutral $B$-meson mixing, i.e.\ $|\Delta B| = 2$.
Again, the \sm{} Wilson coefficient is known through \nlo{}~\cite{Buchalla:1995vs}.
For non-leptonic $|\Delta F| = 1$ decays, the Wilson coefficients $C^\ren_m$ in the \cmm\ basis for the \sm{} can be found in \citeres{Bobeth:1999mk,Gorbahn:2004my} through \nnlo{}.
However, we did not consider the subdominant penguin contributions in our calculation of $\zeta^{-1}(t)$ above.



\section{Conclusions and outlook}

In this contribution we discussed our calculation of the matching matrix of the current-current operators of the electroweak effective Hamiltonian to their flowed counterparts through \nnlo\ \qcd{} published in \citere{Harlander:2022tgk}.
Once the matrix elements from the lattice become available, our results can directly be applied to $K$- or $B$-meson mixing, for example.
The inclusion of penguin operators for non-leptonic $|\Delta F| = 1$ decays is work in progress.
It remains to be seen how the \gff\ approach to flavor physics compares to conventional calculations.

\acknowledgments

I thank Robert Harlander for the collaboration on this project and comments on the manuscript.
Furthermore, I acknowledge financial support by the \textit{Deutsche Forschungsgemeinschaft} (DFG, German Research Foundation) through grant \href{http://gepris.dfg.de/gepris/projekt/386986591?language=en}{386986591} in the early stages of this project and through the Collaborative Research Centre \href{http://p3h.particle.kit.edu/start}{TRR 257} funded through grant \href{http://gepris.dfg.de/gepris/projekt/396021762?language=en}{396021762}.


\bibliographystyle{JHEP}
\bibliography{bib}

\providecommand{\href}[2]{#2}\begingroup\raggedright\begin{thebibliography}{10}

\bibitem{Narayanan:2006rf}
R.~Narayanan and H.~Neuberger, \emph{{Infinite N phase transitions in continuum
  Wilson loop operators}},
  \href{https://doi.org/10.1088/1126-6708/2006/03/064}{\emph{JHEP} {\bfseries
  03} (2006) 064} [\href{https://arxiv.org/abs/hep-th/0601210}{{\ttfamily
  hep-th/0601210}}].

\bibitem{Luscher:2009eq}
M.~L{\"u}scher, \emph{{Trivializing Maps, the Wilson Flow and the HMC
  Algorithm}}, \href{https://doi.org/10.1007/s00220-009-0953-7}{\emph{Commun.
  Math. Phys.} {\bfseries 293} (2010) 899}
  [\href{https://arxiv.org/abs/0907.5491}{{\ttfamily 0907.5491}}].

\bibitem{Luscher:2010iy}
M.~L\"uscher, \emph{{Properties and uses of the Wilson flow in lattice QCD}},
  \href{https://doi.org/10.1007/JHEP08(2010)071}{\emph{JHEP} {\bfseries 08}
  (2010) 071} [\href{https://arxiv.org/abs/1006.4518}{{\ttfamily 1006.4518}}].

\bibitem{Suzuki:2013gza}
H.~Suzuki, \emph{{Energy\textendash{}momentum tensor from the
  Yang\textendash{}Mills gradient flow}},
  \href{https://doi.org/10.1093/ptep/ptt059}{\emph{PTEP} {\bfseries 2013}
  (2013) 083B03} [\href{https://arxiv.org/abs/1304.0533}{{\ttfamily
  1304.0533}}].

\bibitem{Makino:2014taa}
H.~Makino and H.~Suzuki, \emph{{Lattice energy\textendash{}momentum tensor from
  the Yang\textendash{}Mills gradient flow\textemdash{}inclusion of fermion
  fields}}, \href{https://doi.org/10.1093/ptep/ptu070}{\emph{PTEP} {\bfseries
  2014} (2014) 063B02} [\href{https://arxiv.org/abs/1403.4772}{{\ttfamily
  1403.4772}}].

\bibitem{Harlander:2018zpi}
R.V.~Harlander, Y.~Kluth and F.~Lange, \emph{{The two-loop
  energy\textendash{}momentum tensor within the gradient-flow formalism}},
  \href{https://doi.org/10.1140/epjc/s10052-018-6415-7}{\emph{Eur. Phys. J. C}
  {\bfseries 78} (2018) 944}
  [\href{https://arxiv.org/abs/1808.09837}{{\ttfamily 1808.09837}}].

\bibitem{Iritani:2018idk}
T.~Iritani, M.~Kitazawa, H.~Suzuki and H.~Takaura, \emph{{Thermodynamics in
  quenched QCD: energy\textendash{}momentum tensor with two-loop order
  coefficients in the gradient-flow formalism}},
  \href{https://doi.org/10.1093/ptep/ptz001}{\emph{PTEP} {\bfseries 2019}
  (2019) 023B02} [\href{https://arxiv.org/abs/1812.06444}{{\ttfamily
  1812.06444}}].

\bibitem{Taniguchi:2020mgg}
{\scshape WHOT-QCD} collaboration, \emph{{$N_f$ = 2+1 QCD thermodynamics with
  gradient flow using two-loop matching coefficients}},
  \href{https://doi.org/10.1103/PhysRevD.102.014510}{\emph{Phys. Rev. D}
  {\bfseries 102} (2020) 014510}
  [\href{https://arxiv.org/abs/2005.00251}{{\ttfamily 2005.00251}}].

\bibitem{Shirogane:2020muc}
{\scshape WHOT-QCD} collaboration, \emph{{Latent heat and pressure gap at the
  first-order deconfining phase transition of SU(3) Yang-Mills theory using the
  small flow-time expansion method}},
  \href{https://doi.org/10.1093/ptep/ptaa184}{\emph{PTEP} {\bfseries 2021}
  (2021) 013B08} [\href{https://arxiv.org/abs/2011.10292}{{\ttfamily
  2011.10292}}].

\bibitem{Suzuki:2021tlr}
H.~Suzuki and H.~Takaura, \emph{{$t \to 0$ extrapolation function in the small
  flow time expansion method for the energy\textendash{}momentum tensor}},
  \href{https://doi.org/10.1093/ptep/ptab068}{\emph{PTEP} {\bfseries 2021}
  (2021) 073B02} [\href{https://arxiv.org/abs/2102.02174}{{\ttfamily
  2102.02174}}].

\bibitem{Rizik:2020naq}
{\scshape SymLat} collaboration, \emph{{Short flow-time coefficients of
  $CP$-violating operators}},
  \href{https://doi.org/10.1103/PhysRevD.102.034509}{\emph{Phys. Rev. D}
  {\bfseries 102} (2020) 034509}
  [\href{https://arxiv.org/abs/2005.04199}{{\ttfamily 2005.04199}}].

\bibitem{Mereghetti:2021nkt}
E.~Mereghetti, C.J.~Monahan, M.D.~Rizik, A.~Shindler and P.~Stoffer,
  \emph{{One-loop matching for quark dipole operators in a gradient-flow
  scheme}}, \href{https://doi.org/10.1007/JHEP04(2022)050}{\emph{JHEP}
  {\bfseries 04} (2022) 050}
  [\href{https://arxiv.org/abs/2111.11449}{{\ttfamily 2111.11449}}].

\bibitem{Harlander:2020duo}
R.V.~Harlander, F.~Lange and T.~Neumann, \emph{{Hadronic vacuum polarization
  using gradient flow}},
  \href{https://doi.org/10.1007/JHEP08(2020)109}{\emph{JHEP} {\bfseries 08}
  (2020) 109} [\href{https://arxiv.org/abs/2007.01057}{{\ttfamily
  2007.01057}}].

\bibitem{Suzuki:2020zue}
A.~Suzuki, Y.~Taniguchi, H.~Suzuki and K.~Kanaya, \emph{{Four quark operators
  for kaon bag parameter with gradient flow}},
  \href{https://doi.org/10.1103/PhysRevD.102.034508}{\emph{Phys. Rev. D}
  {\bfseries 102} (2020) 034508}
  [\href{https://arxiv.org/abs/2006.06999}{{\ttfamily 2006.06999}}].

\bibitem{Harlander:2021jcq}
R.V.~Harlander and F.~Lange, \emph{{The electroweak Hamiltonian in the gradient
  flow formalism}}, \href{https://doi.org/10.22323/1.398.0415}{\emph{PoS}
  {\bfseries EPS-HEP2021} (2022) 415}
  [\href{https://arxiv.org/abs/2110.15759}{{\ttfamily 2110.15759}}].

\bibitem{Harlander:2022tgk}
R.V.~Harlander and F.~Lange, \emph{{Effective electroweak Hamiltonian in the
  gradient-flow formalism}},
  \href{https://doi.org/10.1103/PhysRevD.105.L071504}{\emph{Phys. Rev. D}
  {\bfseries 105} (2022) L071504}
  [\href{https://arxiv.org/abs/2201.08618}{{\ttfamily 2201.08618}}].

\bibitem{Chetyrkin:1997gb}
K.~Chetyrkin, M.~Misiak and M.~M{\"u}nz, \emph{{$||\Delta F|| = 1$ non-leptonic
  effective Hamiltonian in a simpler scheme}},
  \href{https://doi.org/10.1016/S0550-3213(98)00131-X}{\emph{Nucl. Phys. B}
  {\bfseries 520} (1998) 279}
  [\href{https://arxiv.org/abs/hep-ph/9711280}{{\ttfamily hep-ph/9711280}}].

\bibitem{Buras:1989xd}
A.J.~Buras and P.H.~Weisz, \emph{{QCD nonleading corrections to weak decays in
  dimensional regularization and 't Hooft-Veltman schemes}},
  \href{https://doi.org/10.1016/0550-3213(90)90223-Z}{\emph{Nucl. Phys. B}
  {\bfseries 333} (1990) 66}.

\bibitem{Luscher:2013cpa}
M.~L{\"u}scher, \emph{{Chiral symmetry and the Yang-Mills gradient flow}},
  \href{https://doi.org/10.1007/JHEP04(2013)123}{\emph{JHEP} {\bfseries 04}
  (2013) 123} [\href{https://arxiv.org/abs/1302.5246}{{\ttfamily 1302.5246}}].

\bibitem{Luscher:2011bx}
M.~L{\"u}scher and P.~Weisz, \emph{{Perturbative analysis of the gradient flow
  in non-abelian gauge theories}},
  \href{https://doi.org/10.1007/JHEP02(2011)051}{\emph{JHEP} {\bfseries 02}
  (2011) 051} [\href{https://arxiv.org/abs/1101.0963}{{\ttfamily 1101.0963}}].

\bibitem{Artz:2019bpr}
J.~Artz, R.V.~Harlander, F.~Lange, T.~Neumann and M.~Prausa, \emph{{Results and
  techniques for higher order calculations within the gradient-flow
  formalism}}, \href{https://doi.org/10.1007/JHEP06(2019)121}{\emph{JHEP}
  {\bfseries 06} (2019) 121}
  [\href{https://arxiv.org/abs/1905.00882}{{\ttfamily 1905.00882}}].

\bibitem{Dugan:1990df}
M.J.~Dugan and B.~Grinstein, \emph{{On the vanishing of evanescent operators}},
  \href{https://doi.org/10.1016/0370-2693(91)90680-O}{\emph{Phys. Lett. B}
  {\bfseries 256} (1991) 239}.

\bibitem{Herrlich:1994kh}
S.~Herrlich and U.~Nierste, \emph{{Evanescent operators, scheme dependences and
  double insertions}},
  \href{https://doi.org/10.1016/0550-3213(95)00474-7}{\emph{Nucl. Phys. B}
  {\bfseries 455} (1995) 39}
  [\href{https://arxiv.org/abs/hep-ph/9412375}{{\ttfamily hep-ph/9412375}}].

\bibitem{Gorishnii:1983su}
S.G.~Gorishny, S.A.~Larin and F.V.~Tkachov, \emph{{The algorithm for OPE
  coefficient functions in the MS scheme}},
  \href{https://doi.org/10.1016/0370-2693(83)91439-9}{\emph{Phys. Lett. B}
  {\bfseries 124} (1983) 217}.

\bibitem{Gorishnii:1986gn}
S.G.~Gorishny and S.A.~Larin, \emph{{Coefficient functions of asymptotic
  operator expansions in the minimal subtraction scheme}},
  \href{https://doi.org/10.1016/0550-3213(87)90283-5}{\emph{Nucl. Phys. B}
  {\bfseries 283} (1987) 452}.

\bibitem{Harlander:2020cyh}
R.V.~Harlander, S.Y.~Klein and M.~Lipp, \emph{{FeynGame}},
  \href{https://doi.org/10.1016/j.cpc.2020.107465}{\emph{Comput. Phys. Commun.}
  {\bfseries 256} (2020) 107465}
  [\href{https://arxiv.org/abs/2003.00896}{{\ttfamily 2003.00896}}].

\bibitem{Harlander:1998cmq}
R.~Harlander, T.~Seidensticker and M.~Steinhauser, \emph{{Complete corrections
  of $O(\alpha \alpha_s)$ to the decay of the Z boson into bottom quarks}},
  \href{https://doi.org/10.1016/S0370-2693(98)00220-2}{\emph{Phys. Lett. B}
  {\bfseries 426} (1998) 125}
  [\href{https://arxiv.org/abs/hep-ph/9712228}{{\ttfamily hep-ph/9712228}}].

\bibitem{Seidensticker:1999bb}
T.~Seidensticker, \emph{{Automatic application of successive asymptotic
  expansions of Feynman diagrams}},  in \emph{{6th Conference of the ACAT
  series}}, 1999 [\href{https://arxiv.org/abs/hep-ph/9905298}{{\ttfamily
  hep-ph/9905298}}].

\bibitem{Nogueira:1991ex}
P.~Nogueira, \emph{{Automatic Feynman Graph Generation}},
  \href{https://doi.org/10.1006/jcph.1993.1074}{\emph{J. Comput. Phys.}
  {\bfseries 105} (1993) 279}.

\bibitem{Nogueira:2006pq}
P.~Nogueira, \emph{{Abusing qgraf}},
  \href{https://doi.org/10.1016/j.nima.2005.11.151}{\emph{Nucl. Instrum. Meth.
  A} {\bfseries 559} (2006) 220}.

\bibitem{Vermaseren:2000nd}
J.A.M.~Vermaseren, \emph{{New features of FORM}},
  \href{https://arxiv.org/abs/math-ph/0010025}{{\ttfamily math-ph/0010025}}.

\bibitem{Kuipers:2012rf}
J.~Kuipers, T.~Ueda, J.A.M.~Vermaseren and J.~Vollinga, \emph{{FORM version
  4.0}}, \href{https://doi.org/10.1016/j.cpc.2012.12.028}{\emph{Comput. Phys.
  Commun.} {\bfseries 184} (2013) 1453}
  [\href{https://arxiv.org/abs/1203.6543}{{\ttfamily 1203.6543}}].

\bibitem{vanRitbergen:1998pn}
T.~van Ritbergen, A.N.~Schellekens and J.A.M.~Vermaseren, \emph{{Group theory
  factors for Feynman diagrams}},
  \href{https://doi.org/10.1142/S0217751X99000038}{\emph{Int. J. Mod. Phys. A}
  {\bfseries 14} (1999) 41}
  [\href{https://arxiv.org/abs/hep-ph/9802376}{{\ttfamily hep-ph/9802376}}].

\bibitem{Maierhofer:2017gsa}
P.~Maierh\"ofer, J.~Usovitsch and P.~Uwer, \emph{{Kira\textemdash{}A Feynman
  integral reduction program}},
  \href{https://doi.org/10.1016/j.cpc.2018.04.012}{\emph{Comput. Phys. Commun.}
  {\bfseries 230} (2018) 99}
  [\href{https://arxiv.org/abs/1705.05610}{{\ttfamily 1705.05610}}].

\bibitem{Klappert:2020nbg}
J.~Klappert, F.~Lange, P.~Maierh\"ofer and J.~Usovitsch, \emph{{Integral
  reduction with Kira 2.0 and finite field methods}},
  \href{https://doi.org/10.1016/j.cpc.2021.108024}{\emph{Comput. Phys. Commun.}
  {\bfseries 266} (2021) 108024}
  [\href{https://arxiv.org/abs/2008.06494}{{\ttfamily 2008.06494}}].

\bibitem{Klappert:2019emp}
J.~Klappert and F.~Lange, \emph{{Reconstructing rational functions with
  FireFly}}, \href{https://doi.org/10.1016/j.cpc.2019.106951}{\emph{Comput.
  Phys. Commun.} {\bfseries 247} (2020) 106951}
  [\href{https://arxiv.org/abs/1904.00009}{{\ttfamily 1904.00009}}].

\bibitem{Klappert:2020aqs}
J.~Klappert, S.Y.~Klein and F.~Lange, \emph{{Interpolation of dense and sparse
  rational functions and other improvements in FireFly}},
  \href{https://doi.org/10.1016/j.cpc.2021.107968}{\emph{Comput. Phys. Commun.}
  {\bfseries 264} (2021) 107968}
  [\href{https://arxiv.org/abs/2004.01463}{{\ttfamily 2004.01463}}].

\bibitem{Gambino:2003zm}
P.~Gambino, M.~Gorbahn and U.~Haisch, \emph{{Anomalous dimension matrix for
  radiative and rare semileptonic B decays up to three loops}},
  \href{https://doi.org/10.1016/j.nuclphysb.2003.09.024}{\emph{Nucl. Phys. B}
  {\bfseries 673} (2003) 238}
  [\href{https://arxiv.org/abs/hep-ph/0306079}{{\ttfamily hep-ph/0306079}}].

\bibitem{Gorbahn:2004my}
M.~Gorbahn and U.~Haisch, \emph{{Effective Hamiltonian for non-leptonic
  $|\Delta F| = 1$ decays at NNLO in QCD}},
  \href{https://doi.org/10.1016/j.nuclphysb.2005.01.047}{\emph{Nucl. Phys. B}
  {\bfseries 713} (2005) 291}
  [\href{https://arxiv.org/abs/hep-ph/0411071}{{\ttfamily hep-ph/0411071}}].

\bibitem{Buras:2006gb}
A.J.~Buras, M.~Gorbahn, U.~Haisch and U.~Nierste, \emph{{Charm quark
  contribution to $K^+ \to \pi^+ \nu \bar\nu$ at next-to-next-to-leading
  order}}, \href{https://doi.org/10.1007/JHEP11(2012)167}{\emph{JHEP}
  {\bfseries 11} (2006) 002}
  [\href{https://arxiv.org/abs/hep-ph/0603079}{{\ttfamily hep-ph/0603079}}].

\bibitem{Buchalla:1995vs}
G.~Buchalla, A.J.~Buras and M.E.~Lautenbacher, \emph{{Weak decays beyond
  leading logarithms}},
  \href{https://doi.org/10.1103/RevModPhys.68.1125}{\emph{Rev. Mod. Phys.}
  {\bfseries 68} (1996) 1125}
  [\href{https://arxiv.org/abs/hep-ph/9512380}{{\ttfamily hep-ph/9512380}}].

\bibitem{Brod:2010mj}
J.~Brod and M.~Gorbahn, \emph{{$\epsilon_K$ at next-to-next-to-leading order:
  The charm-top-quark contribution}},
  \href{https://doi.org/10.1103/PhysRevD.82.094026}{\emph{Phys. Rev. D}
  {\bfseries 82} (2010) 094026}
  [\href{https://arxiv.org/abs/1007.0684}{{\ttfamily 1007.0684}}].

\bibitem{Brod:2011ty}
J.~Brod and M.~Gorbahn, \emph{{Next-to-Next-to-Leading-Order Charm-Quark
  Contribution to the $CP$ Violation Parameter $\epsilon_K$ and $\Delta M_K$}},
  \href{https://doi.org/10.1103/PhysRevLett.108.121801}{\emph{Phys. Rev. Lett.}
  {\bfseries 108} (2012) 121801}
  [\href{https://arxiv.org/abs/1108.2036}{{\ttfamily 1108.2036}}].

\bibitem{Bobeth:1999mk}
C.~Bobeth, M.~Misiak and J.~Urban, \emph{{Photonic penguins at two loops and
  $m_t$-dependence of $BR[B \to X_s l^+ l^-]$}},
  \href{https://doi.org/10.1016/S0550-3213(00)00007-9}{\emph{Nucl. Phys. B}
  {\bfseries 574} (2000) 291}
  [\href{https://arxiv.org/abs/hep-ph/9910220}{{\ttfamily hep-ph/9910220}}].

\end{thebibliography}\endgroup

\end{document}